\newcommand{\beq}{\begin{equation}}
\newcommand{\eeq}{\end{equation}}

\documentclass[twocolumn,showpacs,preprintnumbers,amsmath,amssymb]{revtex4}
\usepackage{amsmath}
\usepackage{graphicx}
\usepackage{amsfonts}
\usepackage{amssymb}
\usepackage{epsfig}

\begin{document}

\title{Analytic calculation of energies and wave
functions of the quartic and pure quartic oscillators}

\author{E.~Z. Liverts}
\affiliation{Racah Institute of Physics, The Hebrew University, Jerusalem 91904,
Israel}

\author{V.~B. Mandelzweig}
\affiliation{Racah Institute of Physics, The Hebrew University, Jerusalem 91904,
Israel}

\author{F. Tabakin}
\affiliation{Department of Physics and Astronomy, University of Pittsburgh,
Pittsburgh,
 PA 15260}

\begin{abstract}
\smallskip

Ground state energies and wave functions of quartic and pure
quartic oscillators are calculated by first casting the
Schr\"{o}dinger equation into a nonlinear Riccati form and then
solving that nonlinear equation analytically in the first
iteration of the quasilinearization method (QLM). In the QLM
 the nonlinear differential equation is solved by
approximating the nonlinear terms by a sequence of linear
expressions. The QLM is iterative but not perturbative and gives
stable solutions to nonlinear problems without depending on the
existence of a smallness parameter.  Our explicit analytic results
are then compared with exact numerical and also with WKB solutions
and it is found that our ground state wave functions, using a
range of small to large coupling constants, yield a precision of
between 0.1 and 1 percent and are more accurate than WKB solutions
by two to three orders of magnitude. In addition, our QLM wave
functions are devoid of unphysical turning point singularities and
thus allow one to make analytical estimates of how variation of
the oscillator parameters affects physical systems that can be
described by the quartic and pure quartic oscillators.

\end{abstract}


\maketitle

\section{Introduction}

A basic nonrelativistic quantum mechanics problem is to solve the
Schr\"{o}dinger equation with a potential $V(x)$ that governs
motion of a given physical system. The first two terms of the
power expansion of a one-dimensional, even potential around an
equilibrium position are \beq \frac{g^{2}x^{2}}{2} + \lambda x^4,
\label{eq:ao} \eeq
where $x$ is the deviation from an equilibrium position. The above potential
describes the dynamics of a great many systems that deviate from the
idealized picture of pure harmonic
motion.  When both $g$ and $\lambda$ are nonzero, we call this
potential a ``quartic" or quartic anharmonic  oscillator; whereas,
when $g=0$ with nonzero $\lambda$  it is dubbed a ``pure quartic"
oscillator.  In addition to providing an excellent description of
spectroscopic molecular vibrational data(see Ref.\cite{Laane}
and references therein), the quartic anharmonic
oscillator (\ref{eq:ao}) also serves as a basic tool for checking
various approximate and perturbative methods in quantum mechanics.
Such an application appears in several recent
field theoretical model studies
\cite{Zam,Cas,Muel,Alva,Path,Chil,Chen,Zap,Gil,Jaf,AHS,AAP,Dus}.

It is well known \cite{Ben, Sim} that for the quartic anharmonic
oscillator the perturbation expansion diverges even for small
couplings and becomes completely useless for strong coupling. In
view of this divergence of perturbation theory, we have adopted
\cite{KM2,KMT} the general and very powerful quasilinearization
method (QLM)~\cite{K,BK,VBM1,MT,VBM2},  which although iterative
is not a perturbative method.
 In QLM the $p$-th order solution of a nonlinear differential equation
 with N variables is obtained by first approximating the nonlinear
 aspect by a sequence of linear terms and then iteratively solving the
associated linear equations.
 This iterative process converges to a solution without requiring
  the existence of a smallness parameter. Properties and applications
  of the quasilinearization method were reviewed recently in \cite{VBM3}.

To apply the quasilinearization method, one first casts the
Schr\"{o}dinger equation into the nonlinear Riccati form and then
 solves that nonlinear equation by the QLM iterations.  In a series of
 publications~\cite{KM2,KMT,KMT1,KM3,KM4}, we have shown that for a range of
anharmonic  and other physical potentials (with both weak and
strong couplings), the QLM iterates display very fast quadratic
convergence. Indeed, after just a few QLM iterations, energies and
wave functions are obtained with extremely high accuracy, reaching
20 significant figures for the energy of the sixth iterate even in
the case of very large coupling constants.

Although numerical solutions using either the QLM or direct
numerical solution of the differential equations can be very
accurate, it is important to also provide
  analytic solutions.   Analytic solutions allow one to gauge the role of
different potential parameters, and explore the influence of such
variations on the properties of the quantum system under study.
However, in contrast to the harmonic oscillator, the anharmonic
oscillator cannot be solved analytically, and thus one usually has
to resort to approximations.

The goal of this paper is to obtain and test approximate analytic
solutions for the quartic and pure quartic oscillators using the
explicit analytic equation for the first QLM iterate. We will show
that both energies and wave functions will be represented by
closed analytic expressions with the accuracy of the wave
functions being  between 0.1 and 1 percent for both small and
large coupling constants. Various accurate analytic expressions
for the {\it energies} have already appeared in the literature
based on using convergent, strong coupling expansions generated by
rearrangement of the usual divergent weak coupling expansion
\cite{Jan} or by some variational requirement \cite{Math}.
However,  accurate analytic expressions representing {\it wave
functions} have not hitherto been known.  That result is provided
here.

\section{Main formulae}

The usual WKB substitution $y(x)=\frac{\psi'(x)}{\psi(x)}$
converts the Schr\"{o}dinger equation to the nonlinear Riccati
form
 \beq \frac{dy(x)}{dx}+ \left[ k^2(x)+y^2(x)\right] =0.
\label{eq:weq} \eeq Here $k^2(x)=2\left[ E-V(x)\right] $, where we
use $m=1, \hbar=1$ units.
 The quasilinearization
\cite{VBM1,VBM2,MT,VBM3} of Eq.(\ref{eq:weq}) leads to the
recurrence differential equation \beq \frac{dy_{p}(x)}{dx}+ (2
y_{p-1}(x)) y_{p}(x)=y_{p-1}^2(x)-k^2(x), \label{eq:qeq} \eeq
where $y_{p}(x)$ is the subsequent $p^{th}$ QLM iterate, which
have the same boundary condition as $y(x)$ of Eq.(\ref{eq:weq}).
Note that Eq.(\ref{eq:qeq}) is a linear equation of the form
$\frac{dy_{p}(x)}{dx}+ f(x) y_{p}(x)=q(x),$ with $f(x)=2
y_{p-1}(x)$ and $q(x) =y_{p-1}^2(x)-k^2(x).$

Let us use Eqs.(\ref{eq:qeq}) to estimate the ground state wave
function and energy of the quartic oscillator. Excited states will
be considered elsewhere.

The ground state wave function is nodeless and for an even
potential (\ref{eq:ao}) should therefore be an even function. Its
logarithmic derivative is necessarily odd, and therefore the
boundary condition obviously is $y(0)=0$ and correspondingly
$y_p(0)=0.$

\subsection{ Linear Initial Condition}

 The zero iterate should be based on physical
considerations. Let us consider first an initial guess $y_0(x)=-g
x$. This linear initial condition completely neglects the
anharmonic term containing $\lambda$ compared with the harmonic
term and thus this initial guess is expected to be reasonable only
for relatively small values of $\lambda$.

Solution of the first order linear differential Eq.(\ref{eq:qeq})
with the above zero boundary condition at the origin can always be
found analytically.  For $p=1$ the solution is
\begin{equation}
y_{1}(x)=2\, e^{gx^{2}}\int_{0}^{x}e^{-gs^{2}}(g^{2}s^{2}+\lambda
s^{4}-E_1)\, ds. \label{eq:ig1}
\end{equation}
Integration by parts, yields an expression for $y_{1}(x)$ that
involves the error function ${\rm Erf}(x)\equiv
\frac{2}{\pi}\int_{0}^{x}e^{-t^{2}}dt$:

\begin{eqnarray}
y_{1}(x)&=& \frac{1}{4g^{5/2}}\{ -2x\sqrt{g}\ (2g^{3}+3 \lambda
 + 2 g
\lambda x^{2})  \nonumber   \\
&+ &e^{gx^{2}}\sqrt{\pi}\  {\rm Erf}(x\sqrt{g})\
[2g^{2}(g-2E_1)+3\lambda
]\ \,  \}. \nonumber   \\
 & & \label{eq:ig2}
 \end{eqnarray}


The asymptotic expression ${\rm Erf}(x) \simeq
(1-\frac{e^{-x^2}}{\sqrt{\pi} x})$ for $|x| \rightarrow \infty$,~
indicates that $y_1(x)$ will be exponentially large for very large
$x$ unless the second term in Eq.(\ref{eq:ig2}) is made zero.
Correspondingly, invoking the  condition
$2g^{2}(g-2E_1)+3\lambda=0$ yields  the energy and the logarithmic
derivative in the first iteration: $E_1=\frac{a}{2}$ and
$y_{1}(x)=-ax-bx^{3}$, where $a=g+\frac{3\lambda}{2g^{2}};\,\,\,
b=\frac{\lambda}{g}$. This leads to the first QLM iteration wave
function $\psi(x)=C \exp{\left[-\frac{g x^2}{2}-\lambda
\left(\frac{3 x^2}{4 g^2} +\frac{x^4}{4 g}\right) \right]}$.  This
QLM result for the energy coincides with the perturbative result,
as well as with the result obtained by Friedberg, Lee and Zhao
\cite{FL} who used their recently developed iterative method for
solving the Schr\"{o}dinger equation.

The wave functions we obtained above obviously have incorrect
asymptotic behavior. Also, the energies $E_1$ calculated for
different
 $\lambda,$ as displayed in Table 1, are far from being precise.  Therefore, to improve
 the result one is tempted
to go to the second QLM iteration, using  $y_{1}(x)$ as an input.

Eq.(\ref{eq:qeq}) then yields the second iterate
\begin{eqnarray}\label{ap1}
y_{2}(x)&=&e^{ax^{2}+\frac{b}{2}x^{4}} \int_{0}^{x} \{ [
a^{2}+g^{2}+ 2 ( a b + \lambda )t^2 \\ \nonumber &+& b^{2} t^{4}
]t^{2}-2E_2\} e^{-at^{2}-\frac{b}{2}t^{4}}dt. 
\end{eqnarray}
Since $y_{2}(x)e^{-(ax^{2}+\frac{b}{2}x^{4}})$  approaches a
constant when $x$ goes to infinity, $y_{2}(x)$ and consequently
the corresponding wave function grows exponentially  at infinity,
unless the integral in Eq.(\ref{ap1}) equals  zero when its upper
limit equals infinity. This condition yields the following
expression for the energy $E_2$ in the second iteration:
\begin{equation}
E_2=\frac{\int_{0}^{\infty}\left[a^{2}+g^{2}+2\left(ab+\lambda\right)
t^{2}+b^{2}t^{4}\right]t^{2}e^{-at^{2}-\frac{b}{2}t^{4}}dt}{2\int_{0}
^{\infty}e^{-at^{2}-\frac{b}{2}t^{4}}dt}.
\label{ap2}
\end{equation}
Values of $E_2,$ for this initial linear  form, are compared to
exact values $E$ calculated numerically in Table 1. It is seen
that $E_2$ approximates the exact $E$ reasonably well only for
small $\lambda,$ as we anticipated would result from using an
initial linear condition. We now turn to another choice for the
initial form.

\subsection{ Quadratic Initial Condition}

To ensure a proper wave function asymptotically, one needs
an adequate initial guess.  Our second condition is based on the asymptotic behavior
of the quartic potential.
The zeroth iterate
of the logarithmic derivative, $y_{0}(x),$
is for example now obtained by taking
an initial iterate of quadratic form
$y_{0}(x)=-\sqrt{2 \lambda} \ x^2.$
This choice for the first iterate automatically satisfies the
asymptotic and $x=0$
boundary conditions, but is based on neglecting the harmonic term compare with the anharmonic one.

With this quadratic initial condition, the solution $y_{1}(x)$ of
Eq.(\ref{eq:qeq}) satisfying $y_{1}(0)=0$ is then given by

~\begin{equation}
y_{1}(x)=e^{\frac{2}{3}\sqrt{2 \lambda} \ x^3} \int_{0}^{x}e^{-\frac{2}{3}
\sqrt{2 \lambda} \ s^3}\left[4 \lambda \ s^4+g^2 s^2 -2E_1 \right]ds.
\label{eq:eq1}
\end{equation}
Note, that $y_{1}(x)e^{-\frac{2}{3}\sqrt{2 \lambda}\ x^3}$
approaches a constant, $C_\infty,$ as $x$ goes to infinity, and
consequently $y_{1}(x)$ grows exponentially at infinity unless the
above limit $C_\infty,$
 is set to zero. The latter condition yields another
expression for the energy based on the quadratic initial
condition:

 ~\begin{equation}
 E_1=\frac{\int_{0}^{\infty}e^{-\frac{2}{3}
\sqrt{2 \lambda} \ s^3}\left[2 \lambda \ s^4+\frac{g^2 s^2}{2} \right]ds}
{\int_{0}^{\infty}e^{-\frac{2}{3}
\sqrt{2 \lambda} \ s^3}ds}.
\label{eq:eq2}
\end{equation}

The integrals in Eq.(\ref{eq:eq2}) can be expressed in terms of
the Euler Gamma function $\Gamma(x)$ \cite{GR}. The final
expression for the first iterate energy based on a quadratic
initial condition reads

 ~\begin{equation}
E_1=\frac{\lambda^{\frac{1}{6}}}{3^{\frac{1}{3}}\Gamma\left(\frac{1}{3}\right)}
\left[\frac{3g^{2}}{4\sqrt{\lambda}}+\lambda^{\frac{1}{6}}3^{\frac{2}{3}}
\Gamma\left(\frac{2}{3}\right)\right].
\label{eq:eq3}
\end{equation}

This result should be proper for large $\lambda$ when the second
term of the quartic potential $\frac{g^2 x^2}{2}+\lambda x^4$
dominates over the harmonic term and thus a term containing  $g$
in initial guess $y_0(x)$ could be neglected. The above expression
for the energy is not expected to be suitable for  small
$\lambda$. Indeed, one can see that for the $\lambda \rightarrow
0,$ the energy in Eq.(\ref{eq:eq3}) diverges.

\subsection{Linear Plus Quadratic Initial Condition}

To obtain  a result accommodating arbitrary values of both $g$ and
$\lambda,$ one has to start from an initial choice $y_0(x)$ that
takes into account the asymptotic behavior of both the harmonic
and the anharmonic parts of the potential. Let us choose, for
example, $y_0(x)= -g x - \sqrt{2 \lambda} x^2$ which is a linear
combination of our two previous initial guesses. This yields

\begin{eqnarray}
y_{1}(x)&=&2e^{gx^{2}+\frac{2\sqrt{2\lambda}}{3}x^{3}}
\int_{0}^{x} [ 
t^{2}\left( g^{2}+g\sqrt{2\lambda}t+2\lambda t^{2}\right) 
\nonumber \\
&-&E_1 ] 
e^{-gt^{2}-\frac{2\sqrt{2\lambda}}{3}t^{3}}dt,
\label{eq:eq6}
 \end{eqnarray}
with

~\begin{equation}
E_1=\frac{\int_{0}^{\infty}t^{2}\left(g^{2}+g\sqrt{2\lambda}t+2\lambda t^{2}\right)
e^{-gt^{2}-\frac{2\sqrt{2\lambda}}{3}t^{3}}dt}{
\int_{0}^{\infty}e^{-gt^{2}-\frac{2\sqrt{2\lambda}}{3}t^{3}}dt}.
\label{eq:eq7}
\end{equation}

Another possible initial guess, which also accounts for the
asymptotic behavior of both harmonic and anharmonic
parts of the potential is $y_0(x)= - \sqrt{g^2 x^2 + 2 \lambda x^4}$.
This guess is easy to justify
by assuming that  $y'(x)$ in Eq.(\ref{eq:weq}) at large $x$ becomes
negligible compare with $y^2(x)$ and therefore $y^2(x)$ can be set equal
$k^2(x) = g^2 x^2 + 2 \lambda x^4 -2 E \approx g^2 x^2 + 2 \lambda x^4$
since in this expression we can neglect $E$ for sufficiently large $x$.

The solution of Eq.(\ref{eq:qeq}) using
the initial
condition
 $ y_0(x)= - \sqrt{g^2 x^2 + 2 \lambda
x^4},$ is only slightly more complicated
than in when one uses the initial guess $y_0(x)= -g x - \sqrt{2 \lambda} x^2,$
and is given by the expression
 ~\begin{equation}
y_{1}(x)=2e^{\frac{(g^2 + 2 \lambda x^2)^{\frac{3}{2}}}{3\lambda}}
\int_{0}^{x}e^{-\frac{(g^2 + 2 \lambda t^2)^{\frac{3}{2}}}{3\lambda}}
\left[g^{2}t^{2}+2\lambda t^{4}-E_1\right]dt.
\label{eq:eq4}
\end{equation}

The condition to avoid exponential behavior of the logarithmic
derivative
 at infinity now yields the following expression for the ground state energy
~\begin{equation}
E_1=\frac{\int_{0}^{\infty}t^{2}\left(g^{2}+2\lambda t^{2}\right)
e^{-\frac{(g^2 + 2 \lambda
t^2)^{\frac{3}{2}}}{3\lambda}}dt}{\int_{0}^{\infty} e^{-\frac{(g^2
+ 2 \lambda t^2)^{\frac{3}{2}}}{3\lambda}}dt}. \label{eq:eq5}
\end{equation}

For a  pure quartic oscillator with $g=0,$ both Eqs.(\ref{eq:eq6})
and (\ref{eq:eq4})
 reduce to
 ~\begin{equation}
y_{1}(x)=2e^{\frac{2 \sqrt{2 \lambda} x^3}{3}}
\int_{0}^{x}e^{-\frac{2 \sqrt{2 \lambda} t^3}{3}}
\left[2\lambda t^{4}-E_1\right]dt
\label{eq:eq9}
\end{equation}
with $E_1$ given by
\begin{equation}
E_{1}=\frac{2\lambda\int_{0}^{\infty}e^{-\frac{2\sqrt{2\lambda}}{3}s^{3}}
s^{4}ds}{\int_{0}^{\infty}
e^{-\frac{2\sqrt{2\lambda}}{3}s^{3}}ds}=
\lambda^{\frac{1}{3}}3^{\frac{1}{3}}\frac{\Gamma\left(\frac{2}{3}\right)}
{\Gamma\left(\frac{1}{3}\right)} \simeq 0.729011 \lambda^\frac{1}{3}.
\label{eq:eq8}
\end{equation}
In view of Eq.(\ref{eq:eq8}) $y_1(x)$ of Eq.(\ref{eq:eq9}) can be
expressed in terms of a special function,; namely, the Exponential
Integral \cite{GR}
${\textbf{EI}}_{\mu}(z)=\int_{1}^{\infty}e^{-zt}t^{-\mu}dt$:
\begin{eqnarray}
y_{1}(x)&=& -\sqrt{2\lambda}\,
x^{2}+\frac{2}{3}e^{\frac{2\sqrt{2\lambda}}{3}x^{3}} \{ 
-\sqrt{2\lambda}\,
x^{2}{\textbf{EI}}_{\frac{1}{3}}\left(\frac{2\sqrt{2\lambda}}{3}x^{3}\right)
\nonumber\\
&+&\lambda^{\frac{1}{3}}3^{\frac{1}{3}}\frac{\Gamma (\frac{2}{3})}
{\Gamma (\frac{1}{3} )}x\,{\textbf{EI}}_{\frac{2}{3}}
\left(\frac{2\sqrt{2\lambda}}{3} x^{3}\right) \} .
\label{eq:eq10}
\end{eqnarray}
This expression for the
log derivative yields a first iterate QLM wave function, based on
the initial condition
 $ y_0(x)= - \sqrt{g^2 x^2 + 2 \lambda
x^4},$  that is the main result of our paper.

The exact dependence
of $E_{1}$ on $\lambda$ for the pure quartic oscillator has the
same form, but with a factor of $0.667\,986\,259$ before
$\lambda^\frac{1}{3}$ \cite{Jan}, so that the accuracy of the QLM
prediction for the energy is about 9.1 percent. The WKB energy can
be easily estimated and gives
 $E_{WKB} \simeq 0.546267 \lambda^\frac{1}{3}$, an accuracy of 18.2 percent.

\newpage

\begin{widetext}
\begin{table}
\caption{Ground state energies $E$ for the quartic oscillator
potential $U(x)=g^{2}x^{2}/2+\lambda x^{4}$ with $g=1$.}
\begin{center}
\begin{tabular}{|c|c|c|c|c|c|c|c|c|c|c|c|c|c|}
\hline
{\scriptsize $\lambda$}&
 {\scriptsize $E_{exact}$}&
 {\scriptsize $E_{WKB}$}&
 {\scriptsize $\Delta E_{WKB}$}&
 {\scriptsize $E^{(1)}$}&
 {\scriptsize $\Delta E^{(1)}$}&
 {\scriptsize $E^{(2)};y_{0}=-gx$}&
 {\scriptsize $\Delta E^{(2)}$}&
 {\scriptsize $E^{(3)}$}&
 {\scriptsize $\Delta E^{(3)}$}&
 {\scriptsize $E^{(4)}$}&
 {\scriptsize $\Delta E^{(4)}$}&
 {\scriptsize $E^{(3)};$ $y_{0}=$}&
 {\scriptsize $\Delta E^{(3)}$}\tabularnewline
&
&
&
&
 {\scriptsize $y_{0}=-gx$}&
 {\scriptsize (\%)}&
 {\scriptsize $2^{nd}iteration$}&
 {\scriptsize (\%)}&
 {\scriptsize $y_{0}=-x^{2}\sqrt{2\lambda}$}&
 {\scriptsize (\%)}&
 {\scriptsize $y_{0}=-gx-x^{2}\sqrt{2\lambda}$}&
 {\scriptsize (\%)}&
 {\scriptsize $-\sqrt{g^{2}x^{2}+2\lambda x^{4}}$}&
 {\scriptsize (\%)}\tabularnewline
\hline
{\scriptsize 0}&
 {\scriptsize 1/2}&
 {\scriptsize 1/2}&
 {\scriptsize 0}&
 {\scriptsize 1/2}&
 {\scriptsize 0}&
 {\scriptsize 1/2}&
 {\scriptsize 0}&
&
&
 {\scriptsize 1/2}&
 {\scriptsize 0}&
 {\scriptsize 1/2}&
 {\scriptsize 0}\tabularnewline
\hline
{\scriptsize 0.1}&
 {\scriptsize 0.55915}&
 {\scriptsize 0.53328}&
 {\scriptsize 4.6}&
 {\scriptsize 0.575}&
 {\scriptsize 2.8}&
 {\scriptsize 0.55983}&
 {\scriptsize 0.1}&
 {\scriptsize 0.75658}&
 {\scriptsize 35}&
 {\scriptsize 0.56940}&
 {\scriptsize 1.8}&
 {\scriptsize 0.56149}&
 {\scriptsize 0.4}\tabularnewline
\hline
{\scriptsize 0.3}&
 {\scriptsize 0.63799}&
 {\scriptsize 0.58466}&
 {\scriptsize 8.3}&
 {\scriptsize 0.725}&
 {\scriptsize 13.6}&
 {\scriptsize 0.64869}&
 {\scriptsize 1.7}&
 {\scriptsize 0.77799}&
 {\scriptsize 21.9}&
 {\scriptsize 0.64838}&
 {\scriptsize 1.6}&
 {\scriptsize 0.64705}&
 {\scriptsize 1.4}\tabularnewline
\hline
{\scriptsize 0.5}&
 {\scriptsize 0.69618}&
 {\scriptsize 0.62538}&
 {\scriptsize 10.2}&
 {\scriptsize 0.875}&
 {\scriptsize 25.7}&
 {\scriptsize 0.72728}&
 {\scriptsize 4.4}&
 {\scriptsize 0.82319}&
 {\scriptsize 18.2}&
 {\scriptsize 0.70552}&
 {\scriptsize 1.3}&
 {\scriptsize 0.71126}&
 {\scriptsize 2.2}\tabularnewline
\hline
{\scriptsize 1}&
 {\scriptsize 0.80377}&
 {\scriptsize 0.70420}&
 {\scriptsize 12.4}&
 {\scriptsize 1.25}&
 {\scriptsize 55.5}&
 {\scriptsize 0.91423}&
 {\scriptsize 13.7}&
 {\scriptsize 0.92313}&
 {\scriptsize 14.8}&
 {\scriptsize 0.81138}&
 {\scriptsize 0.95}&
 {\scriptsize 0.83090}&
 {\scriptsize 3.4}\tabularnewline
\hline
{\scriptsize 2}&
 {\scriptsize 0.95157}&
 {\scriptsize 0.81667}&
 {\scriptsize 14.2}&
 {\scriptsize 2}&
 {\scriptsize 110}&
 {\scriptsize 1.2829}&
 {\scriptsize 34.8}&
 {\scriptsize 1.07257}&
 {\scriptsize 12.7}&
 {\scriptsize 0.95853}&
 {\scriptsize 0.73}&
 {\scriptsize 0.99577}&
 {\scriptsize 4.6}\tabularnewline
\hline
{\scriptsize 10}&
 {\scriptsize 1.50497}&
 {\scriptsize 1.25412}&
 {\scriptsize 16.7}&
 {\scriptsize 8}&
&
 {\scriptsize 4.2628}&
 {\scriptsize 183}&
 {\scriptsize 1.6607}&
 {\scriptsize 10.3}&
 {\scriptsize 1.5259}&
 {\scriptsize 1.4}&
 {\scriptsize 1.61085}&
 {\scriptsize 7.0}\tabularnewline
\hline
{\scriptsize 100}&
 {\scriptsize 3.13138}&
 {\scriptsize 2.57181}&
 {\scriptsize 17.9}&
 {\scriptsize 75.5}&
&
&
&
 {\scriptsize 3.4256}&
 {\scriptsize 9.4}&
 {\scriptsize 3.2564}&
 {\scriptsize 4.0}&
 {\scriptsize 3.40039}&
 {\scriptsize 8.5}\tabularnewline
\hline
{\scriptsize 1000}&
 {\scriptsize 6.69422}&
 {\scriptsize 5.47955}&
 {\scriptsize 18.1}&
 {\scriptsize 750.5}&
&
&
&
 {\scriptsize 7.3095}&
 {\scriptsize 9.2 }&
 {\scriptsize 7.1171}&
 {\scriptsize 6.3}&
 {\scriptsize 7.29744}&
 {\scriptsize 9.0}  \tabularnewline
\hline
\end{tabular}
\end{center}
\end{table}
\end{widetext}
\newpage
\section{Results and discussion}

The ground state energies for
the quartic oscillator in the first
QLM approximation for different initial guesses and for  values of
$g=1$ and $\lambda$ between zero and one thousand and their
comparison with the numerically calculated exact and WKB values
are given in Table 1. One can see that the values computed using
explicit equations (\ref{eq:eq7}) and (\ref{eq:eq5}) for the QLM
energy are significantly more
accurate than the WKB values or than values obtained in the first
and second QLM iterations with the initial guess $y_0(x)=-gx$.
They have a precision of 0.4 to 9 percent for values of $\lambda$
varying between 0.1 and 1000, respectively.

However, the main results of our work, are not the expressions for the
energy. As mentioned in the introduction, such expressions were
already given in different forms by others.   Our major results
are the analytic expressions for the wave functions given by
Eqs.(\ref{eq:eq6}) and Eq.(\ref{eq:eq4}),  which are based on
using the first QLM iterate with the initial conditions
$y_0=-gx-\sqrt{2\lambda} x^2$  and
$y_0=-\sqrt{g^2x^2+2\lambda x^4}$, respectively.

\begin{figure}
\begin{center}
\epsfig{file=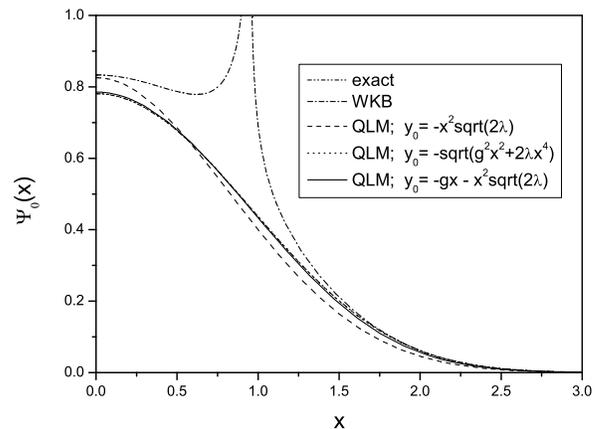,width=87mm}
\end{center}
\caption{Comparison of the WKB, QLM and
exact wave functions for the ground state of the quartic oscillator
 for $g=1, \lambda=0.1$.}
\label{fig1}
\end{figure}
\begin{figure}
\begin{center}
\epsfig{file=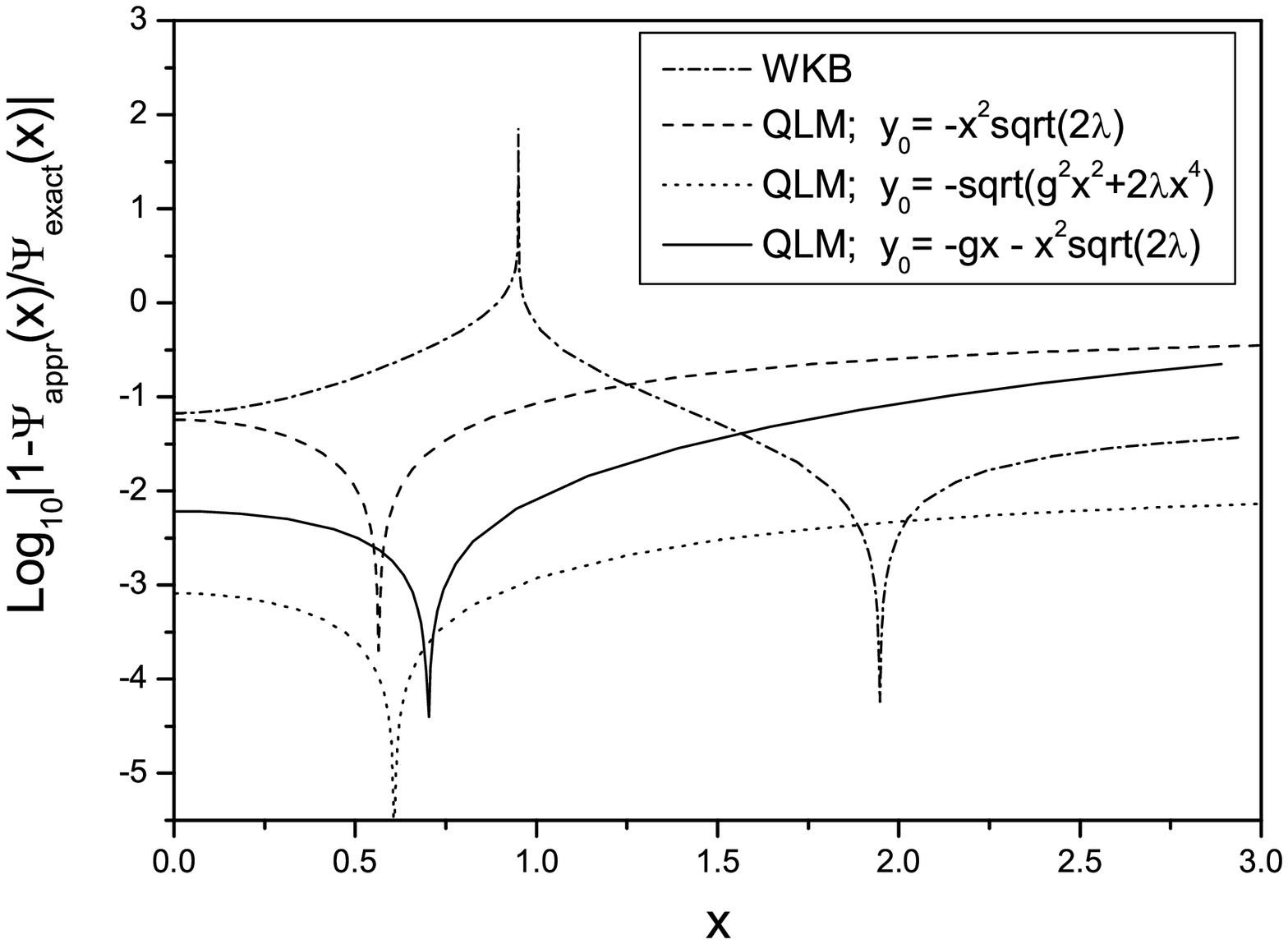,width=87mm}
\end{center}
\caption{Logarithm of the differences of the WKB and QLM wave
functions with exact wave function
for the ground state of the quartic oscillator for $g=1, \lambda=0.1.$}
\label{fig2}
\end{figure}
\begin{figure}
\begin{center}
\epsfig{file=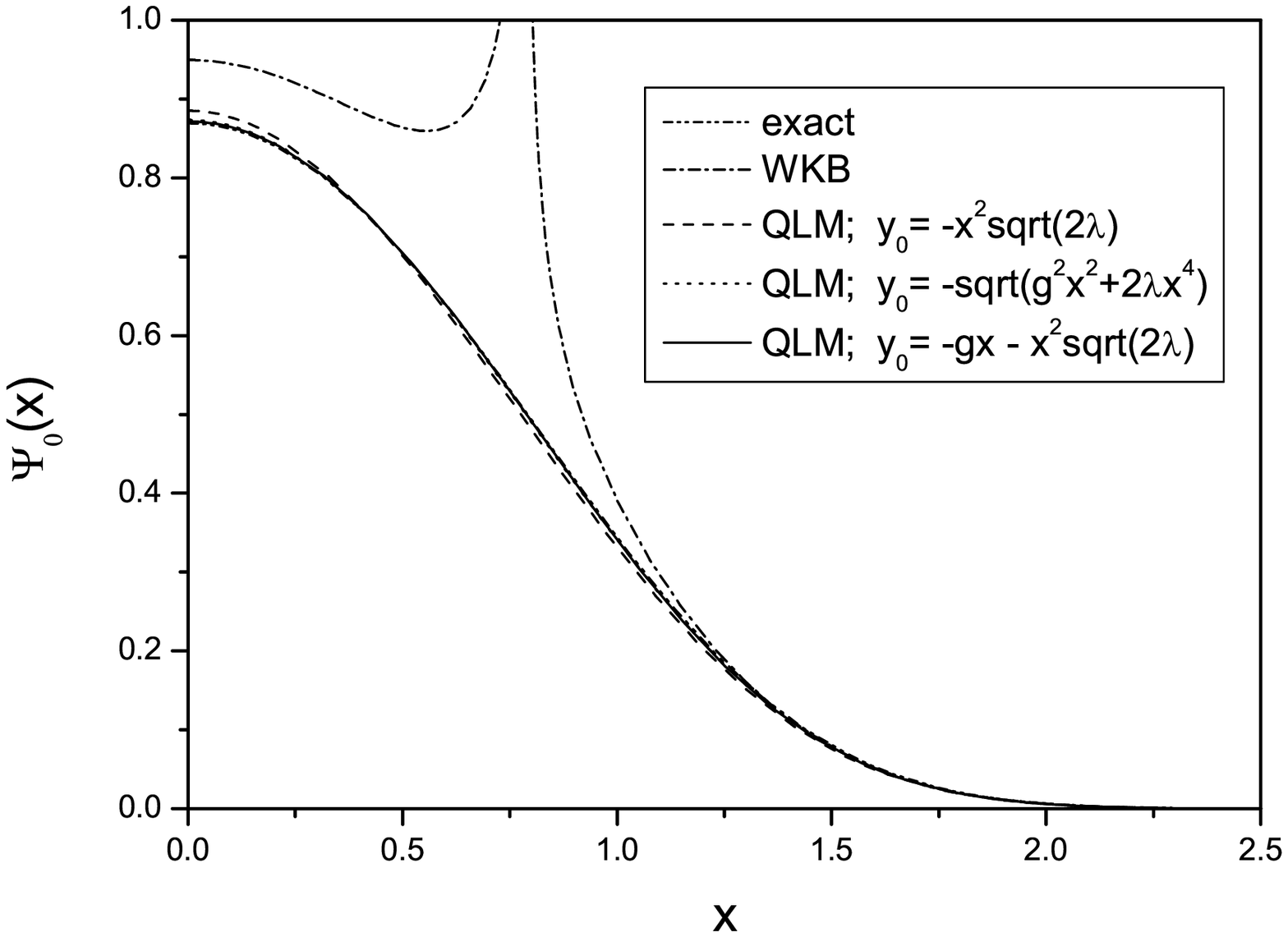,width=87mm}
\end{center}
\caption{Comparison of the WKB, QLM and
exact wave functions for the ground state of the quartic
oscillator for $g=1, \lambda=1$.}
\label{fig3}
\end{figure}
\begin{figure}
\begin{center}
\epsfig{file=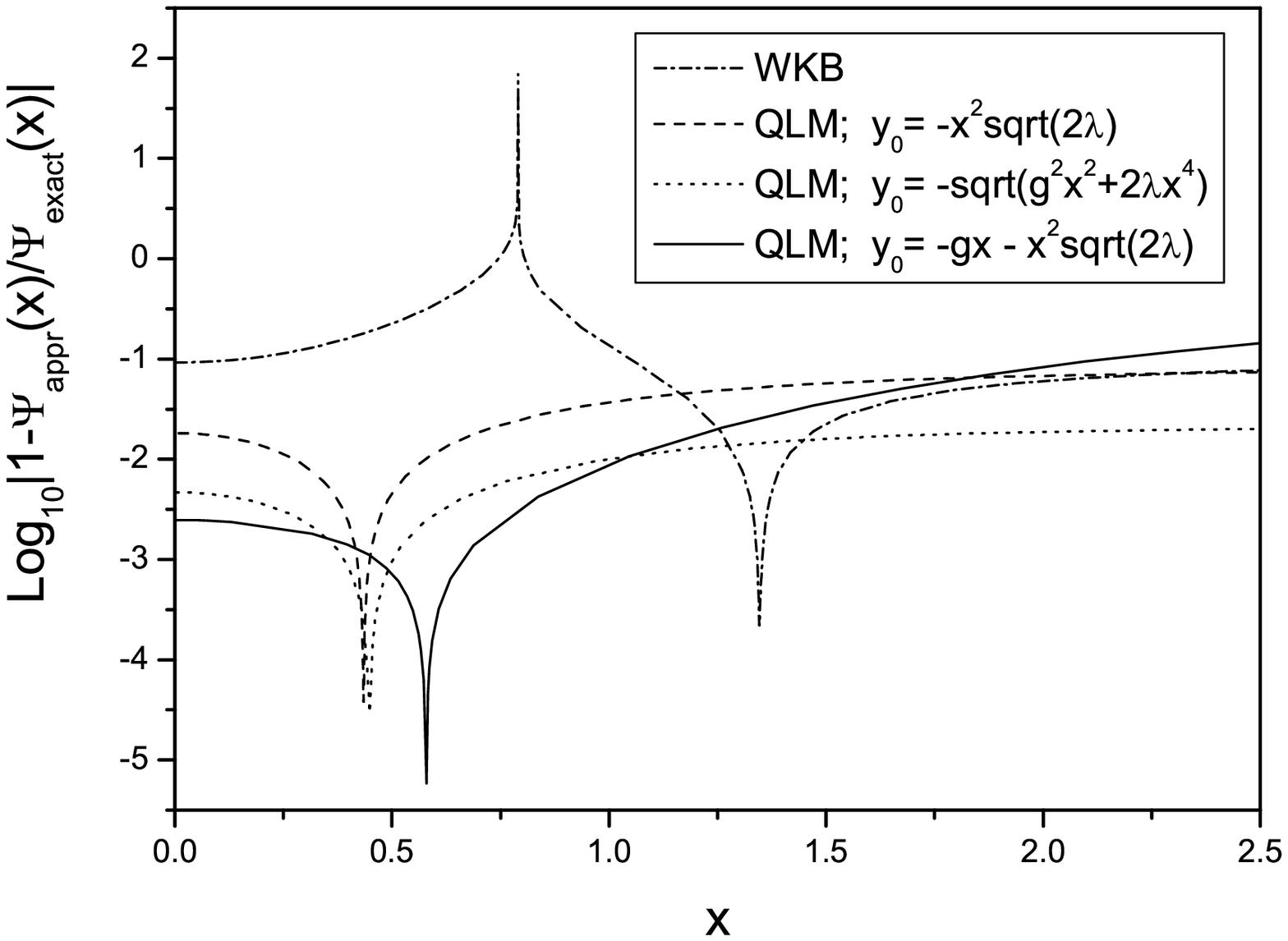,width=87mm}
\end{center}
\caption{Logarithm of the differences of the WKB and QLM wave
functions with exact wave function
for the ground state of the quartic oscillator for $g=1, \lambda=1.$}
\label{fig4}
\end{figure}
\begin{figure}
\begin{center}
\epsfig{file=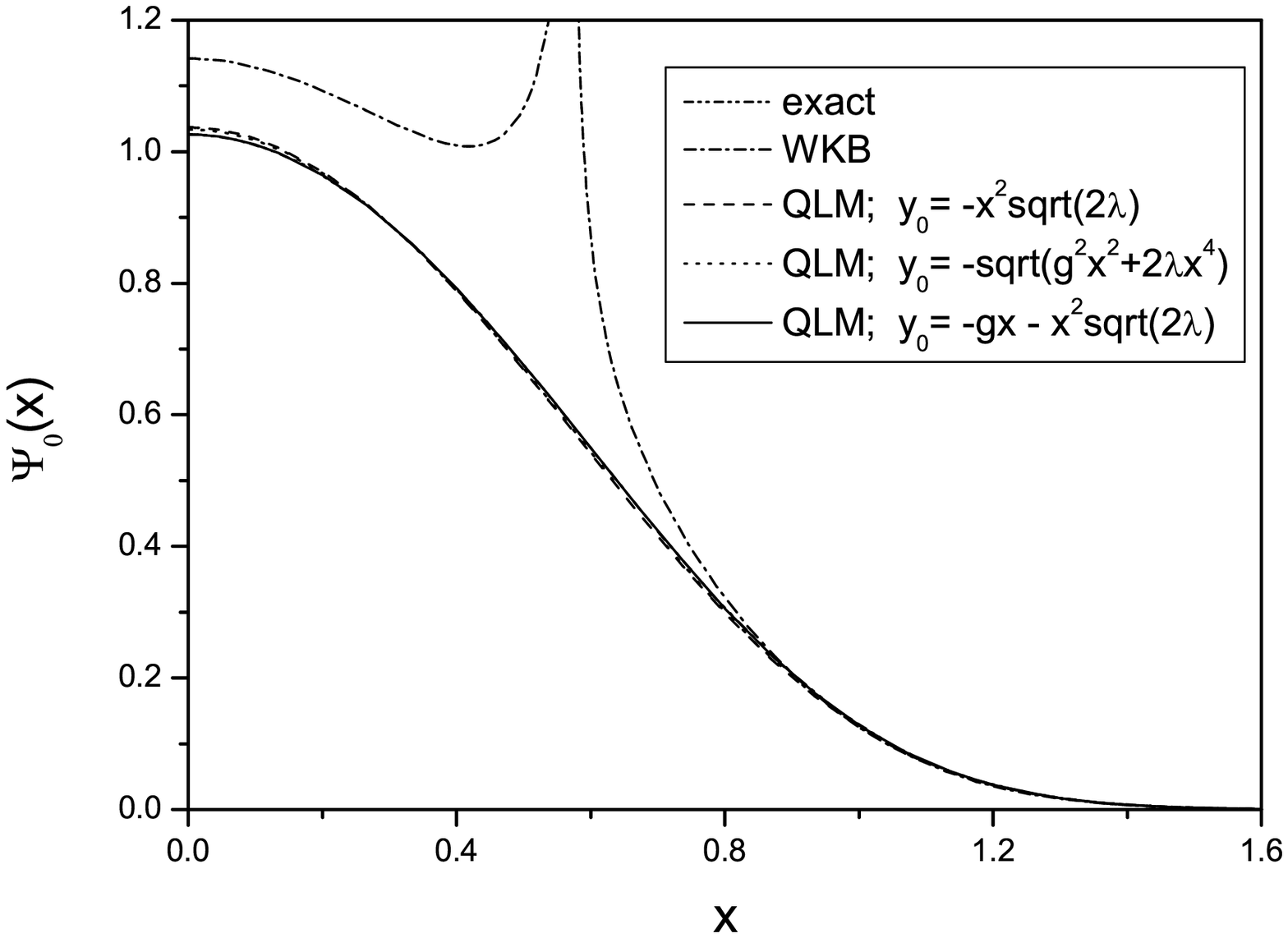,width=87mm}
\end{center}
\caption{Comparison of the WKB, QLM and
exact wave functions for the ground state of the quartic oscillator
 for $g=1, \lambda=10$.}
\label{fig5}
\end{figure}
\begin{figure}
\begin{center}
\epsfig{file=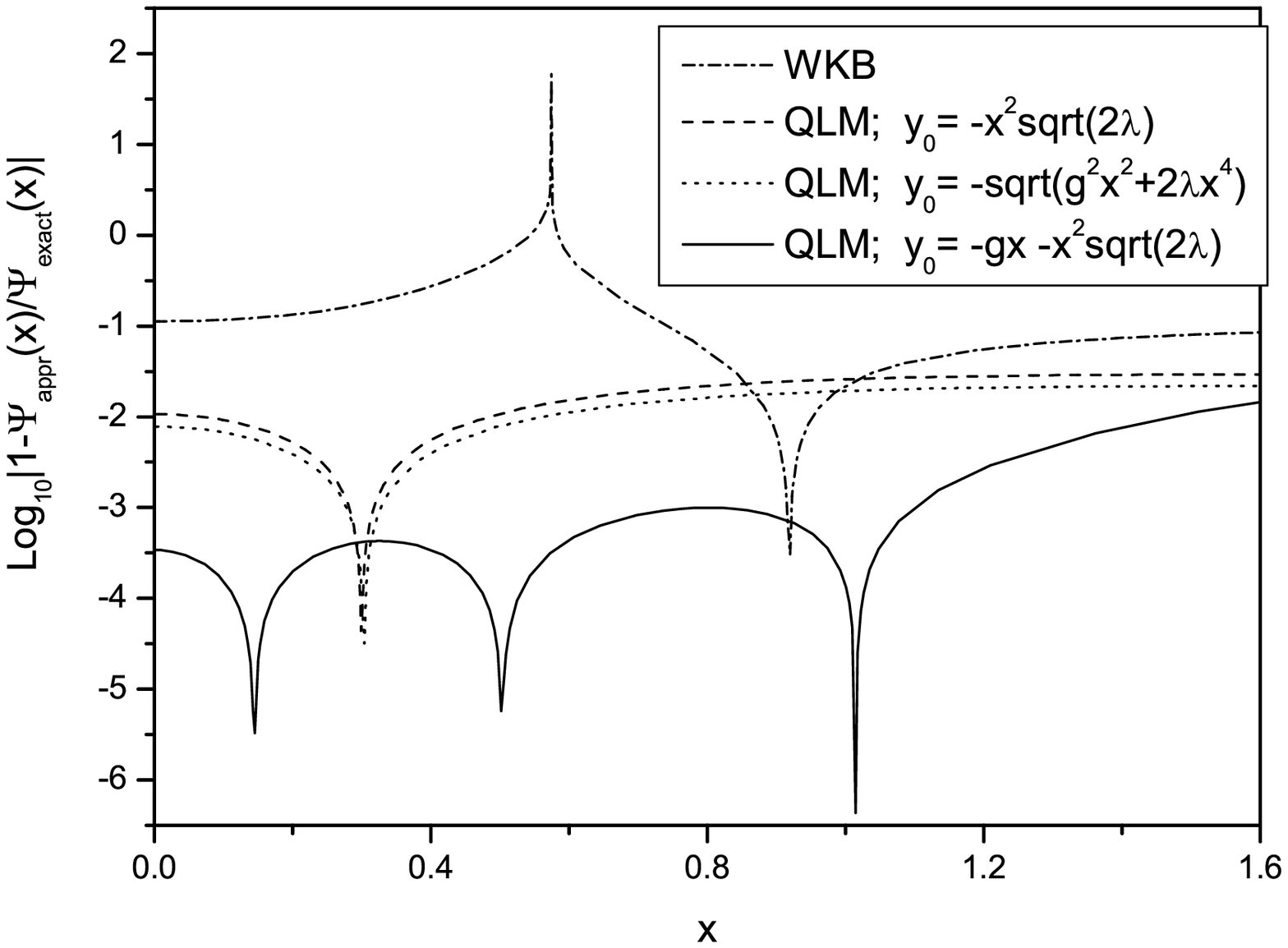,width=87mm}
\end{center}
\caption{Logarithm of the differences of the WKB and QLM wave functions
 with exact wave function
for the ground state of the quartic oscillator for $g=1, \lambda=10.$}
\label{fig6}
\end{figure}
\begin{figure}
\begin{center}
\epsfig{file=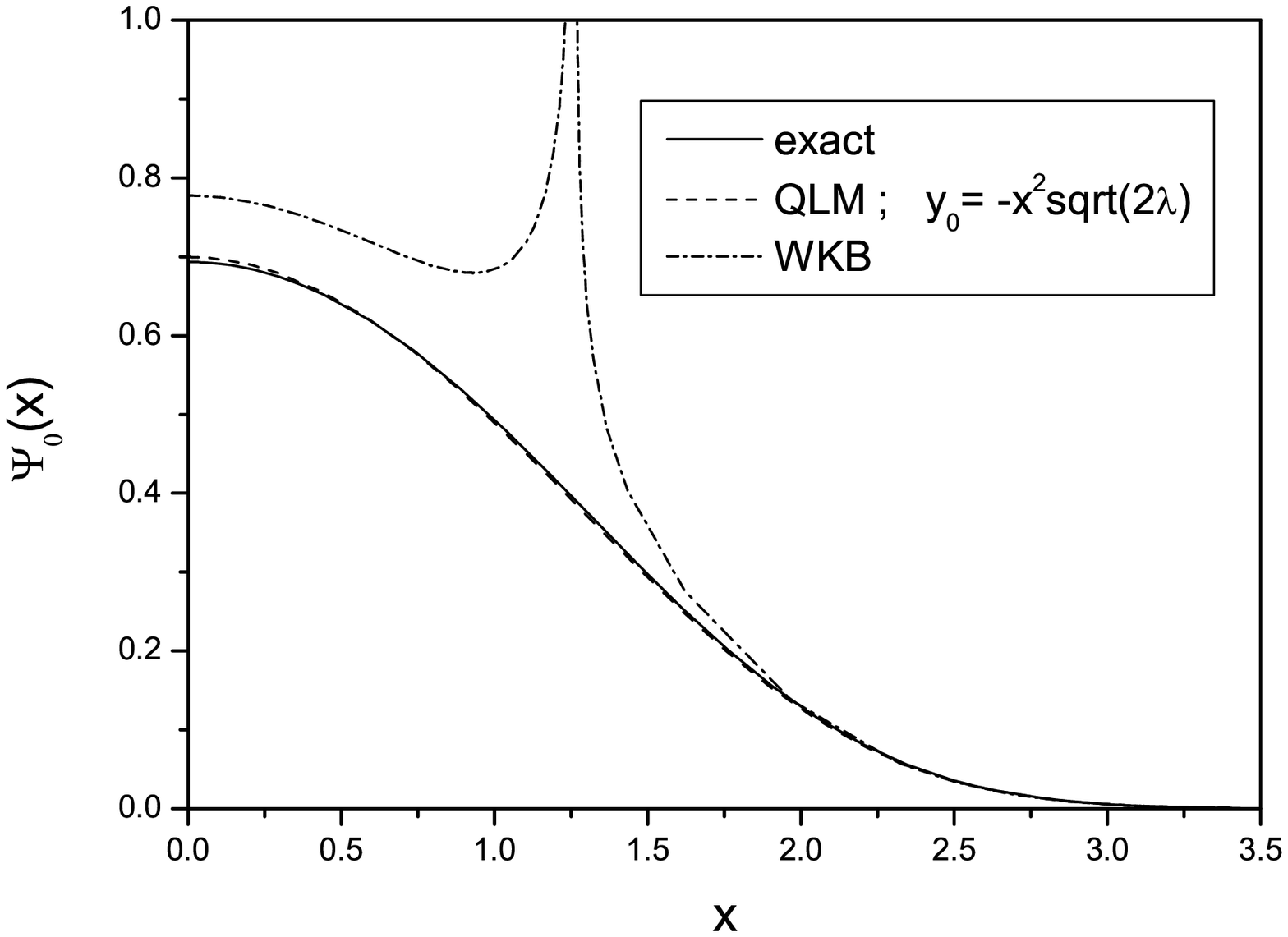,width=87mm}
\end{center}
\caption{Comparison of the WKB, QLM and
exact wave functions for the ground state of the pure quartic
oscillator for $g=0, \lambda=0.1$.}
\label{figp1}
\end{figure}
\begin{figure}
\begin{center}
\epsfig{file=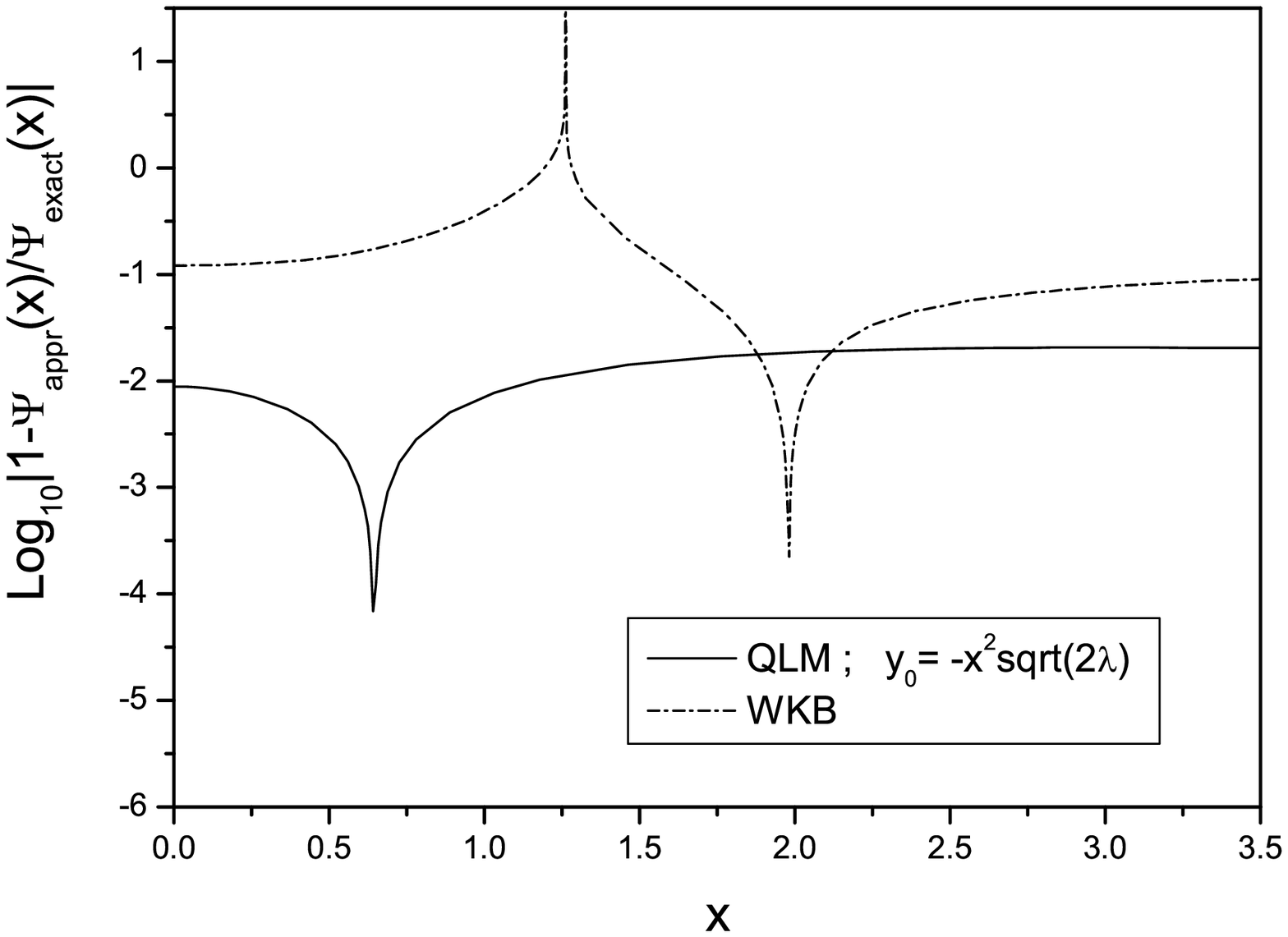,width=87mm}
\end{center}
\caption{Logarithm of the differences of the WKB and QLM wave
functions with exact wave function
for the ground state of the pure quartic  oscillator for $g=0, \lambda=0.1.$}
\label{figp2}
\end{figure}
\begin{figure}
\begin{center}
\epsfig{file=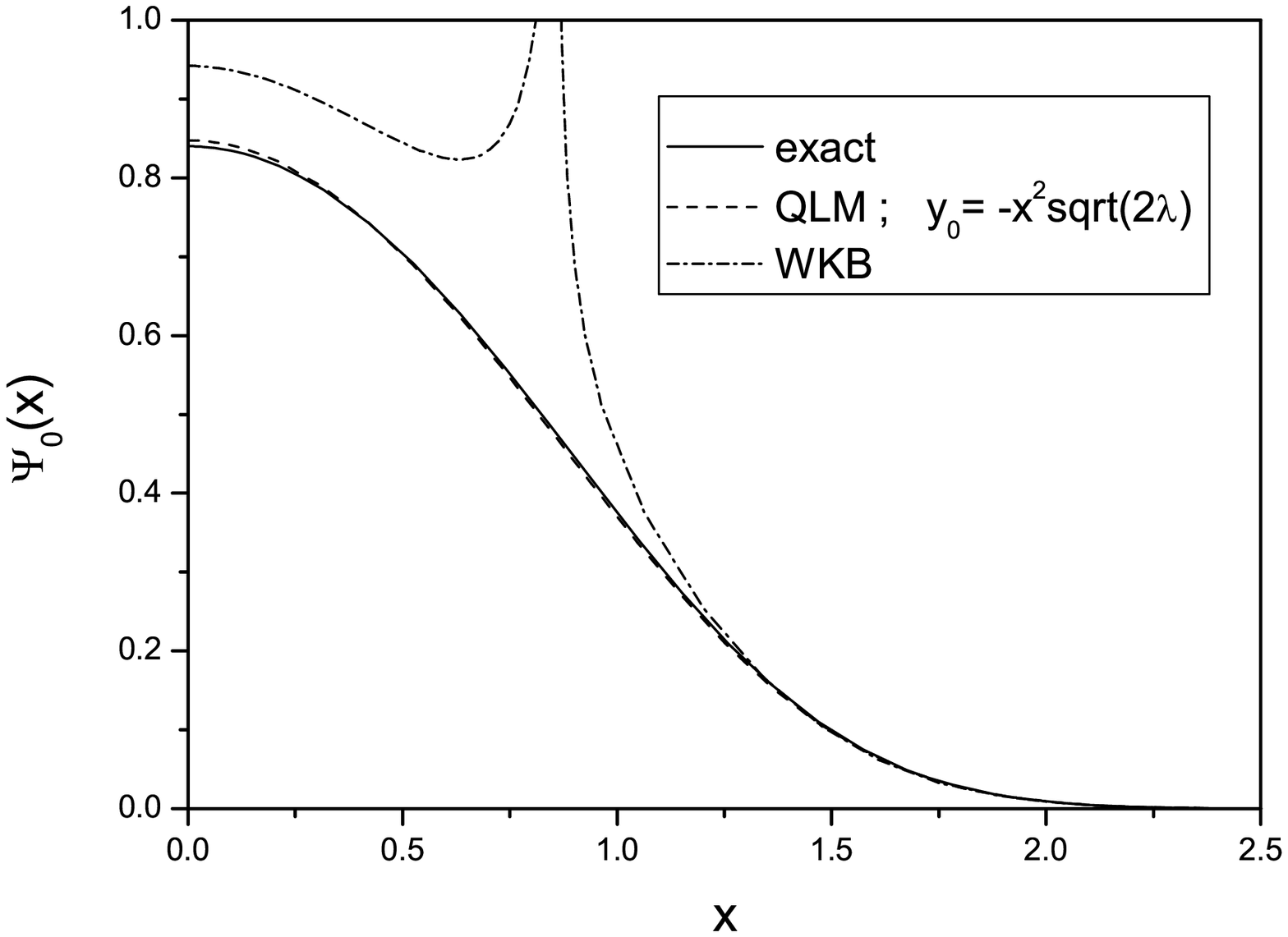,width=87mm}
\end{center}
\caption{Comparison of the WKB, QLM and
exact wave functions for the ground state of the pure quartic
oscillator for $g=0, \lambda=1$.}
\label{figp3}
\end{figure}
\begin{figure}
\begin{center}
\epsfig{file=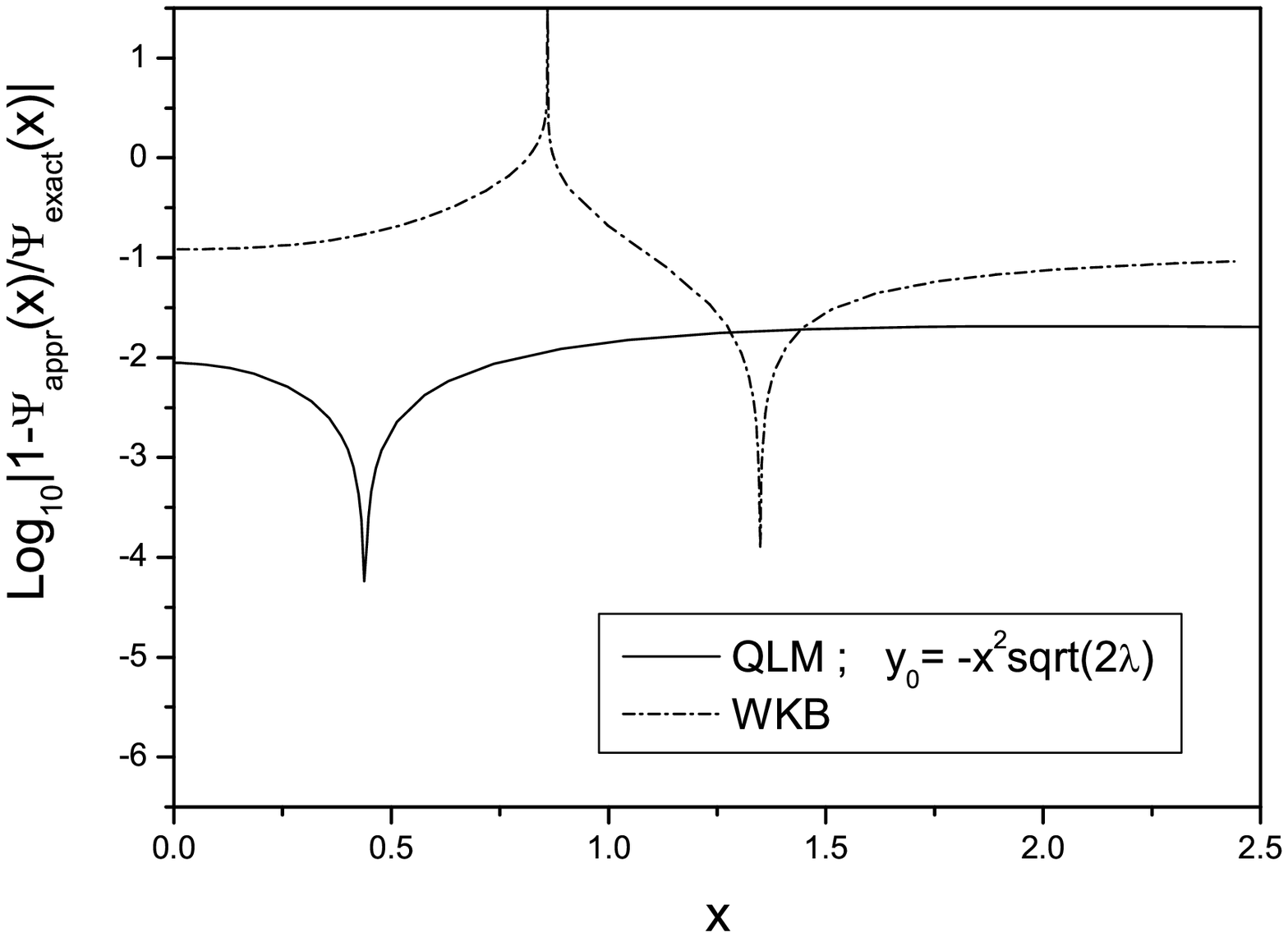,width=87mm}
\end{center}
\caption{Logarithm of the differences of the WKB and QLM wave
functions with exact wave function
for the ground state of the pure quartic oscillator for $g=0, \lambda=1.$}
\label{figx4a}
\end{figure}
\begin{figure}
\begin{center}
\epsfig{file=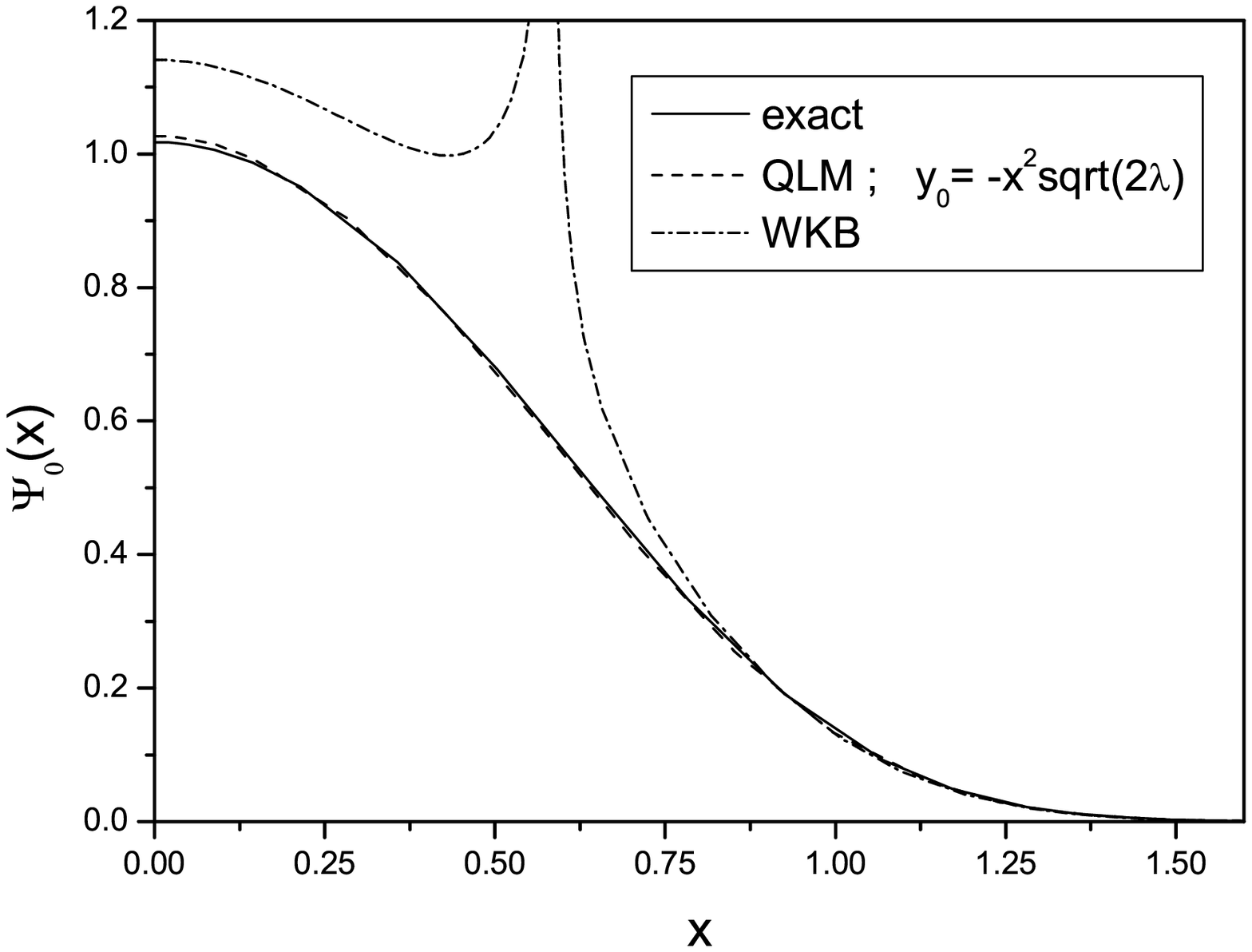,width=87mm}
\end{center}
\caption{Comparison of the WKB, QLM and
exact wave functions for the ground state of the pure quartic
oscillator for $g=0, \lambda=10$.}
\label{figp5}
\end{figure}
\begin{figure}
\begin{center}
\epsfig{file=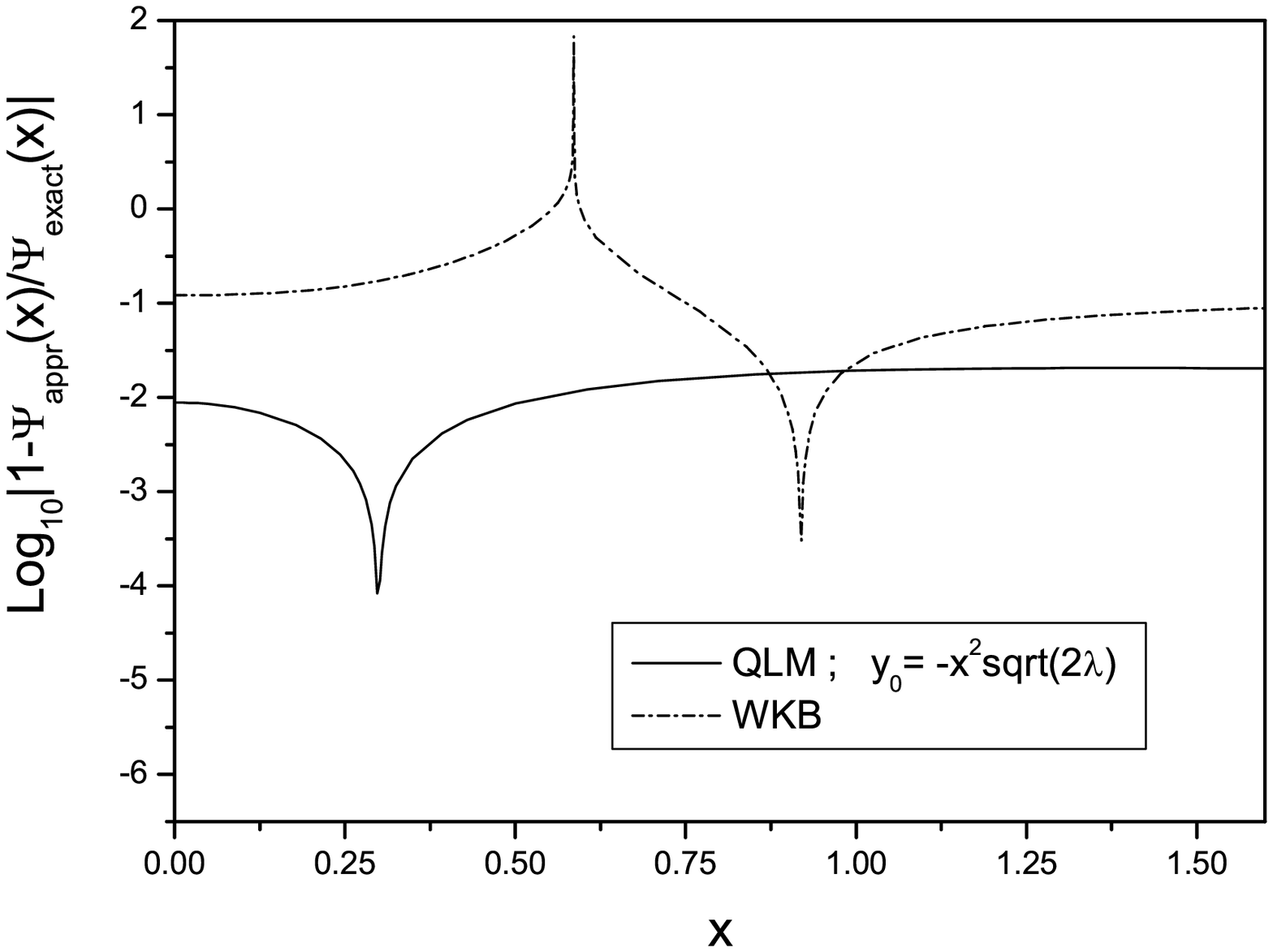,width=87mm}
\end{center}
\caption{Logarithm of the differences of the WKB and QLM wave
 functions with exact wave function
for the ground state of the pure quartic oscillator for $g=0, \lambda=10.$}
\label{figp6}
\end{figure}

The graphs of the wave functions for the quartic oscillator with $g=1$ and
for different $\lambda$ together with the correspondent exact and WKB wave functions,
are presented in Figs. \ref{fig1}, \ref{fig3} and \ref{fig5}, while
Figs. \ref{fig2}, \ref{fig4} and \ref{fig6} display the logarithm 
of the absolute value of the differences
between the WKB or QLM wave functions and the exact solution for
$\lambda$ being equal to 0.1, 1 and 10, respectively. The same
graphs for the pure quartic oscillator ($g=0$)
are presented in Figs. \ref{figp1}, \ref{figp3}, \ref{figp5} and
in Figs. \ref{figp2}, \ref{figx4a}, \ref{figp6}, respectively. One
can see that in all the graphs the differences between the exact
and QLM solutions are two to three orders of magnitude smaller
than the differences between the
exact and the WKB solutions and that the QLM wave functions
expressed analytically by Eqs.(\ref{eq:eq6}),(\ref{eq:eq4}) have
an accuracy of between 0.1 and 1 percent. The order of magnitude
better accuracy of the wave function compared to the poorer
accuracy of the energies is explained by the fact that the general
theorems \cite{K,BK,VBM1,MT,VBM2,VBM3} for the QLM iterates show
that the solutions converge quadratically with each iteration,
while no such  convergence theorem has been proven for the energy
iterates. Note, that the dips in the Figures are artifacts of the
logarithmic scale, since the logarithm of the absolute value of
the difference of two solutions goes to $- \infty$ at points where
the difference changes sign. The overall accuracy of the solution
can therefore be inferred only at $x$ values not too close to the
dips.

\section{Conclusion}

We calculated analytically the ground state energy
and wave function of the quartic and pure quartic oscillators
 by casting the Schr\"{o}dinger equation into the nonlinear Riccati
  form , which is then solved
in the first iteration of the quasilinearization method (QLM),
which approaches the solution of the nonlinear differential
equation  by approximating nonlinear terms with a sequence of
linear ones and does not rely on the existence of a smallness
parameter. Comparison of our results with exact numerical
solutions and the WKB solutions shows that the explicit analytic
expressions we obtain (\ref{eq:eq7}) and  (\ref{eq:eq5}) for the
ground state energy  have a precision of only a few percent while
the analytically expressed wave functions (\ref{eq:eq6}) and
(\ref{eq:eq4}) have an accuracy of between 0.1 and 1 percent and
are more accurate by two to three orders of magnitude  than those
given in  the WKB approximation. The QLM wave function in addition
possess no unphysical turning point singularities which allows one
to use these wave functions to make analytical estimates of the
effects of variation of the oscillator parameters on the
properties of systems described by quartic and pure quartic
oscillators.

The next QLM iterations could be evaluated
numerically \cite{KM2,KMT,KMT1,KM3,KM4}. These further QLM iterates
 for the different anharmonic and other physical potentials with
   both strong and weak couplings also display very fast quadratic
convergence so that the accuracy of energies and wave functions
obtained after a few iterations is extremely high, reaching 20
significant figures for the energy of the sixth iterate even in
the case of very large coupling constants.

Extension of this approach to excited states and to other
potentials is underway.

\newpage

\begin{acknowledgments}
This research was supported by Grant No.\ 2004106 from
the United States-Israel Binational Science Foundation (BSF), Jerusalem, Israel.
\end{acknowledgments}

\end{document}